%% file: fouling_SISSA.tex
\documentclass{elsart_sf}
\usepackage{epsfig,subfigure}
\usepackage{float}
\begin{document}
\begin{frontmatter}
\title{Sedimentation and Fouling of Optical Surfaces at the ANTARES Site}
\vspace{-4mm}
\input{SL_fouling}
\begin{abstract}
ANTARES is a project leading towards the construction and deployment 
of a neutrino telescope in the deep Mediterranean Sea.  The  
telescope  will use an array of photomultiplier tubes to 
detect the Cherenkov light emitted by muons resulting from the 
interaction with matter of high energy neutrinos.  In the vicinity of 
the deployment site the ANTARES collaboration has performed a series 
of in-situ measurements to study the change in light transmission 
through glass surfaces during immersions of several months.  The 
average loss of light transmission is estimated to be only $\sim$~2\% at the
equator of a glass sphere one year after deployment.  It decreases 
with increasing zenith angle, and tends to saturate with time.  The 
transmission loss, therefore, is expected to remain small for the several 
year lifetime of the ANTARES detector whose optical modules are 
oriented downwards.  The measurements were complemented 
by the analysis of the $^{210}$Pb activity profile in sediment cores 
and the study of biofouling on glass plates.  Despite a significant 
sedimentation rate at the site, in the 0.02 -- 0.05~$\rm cm\cdot yr^{-1}$ 
range, the sediments adhere loosely to the glass surfaces and can be 
washed off by water currents.  Further, fouling by deposits of 
light-absorbing particulates is only significant for surfaces facing 
upwards.
\end{abstract}
\vspace{2cm}
\end{frontmatter}

\normalsize
\newpage

\section{Introduction}

The ANTARES undersea neutrino telescope~\cite{WEB,conf} will be dedicated
to the detection of high energy neutrinos. The scientific programme
covers searches for astrophysical, cosmological and dark matter neutrino sources,
as well as the study of flavour oscillations for atmospheric neutrinos.\\
The selection of a suitable site for the telescope takes into account several
parameters. The distance to shore dictates the length of the
electro-optical cable needed to power the detector and transmit the data.
The depth determines the rate of background from down-going cosmic ray
muons, as well as the limiting angle of observation near the horizon.
The nature and topography of the sea floor are relevant for the
anchoring of the detector elements.  Deep sea currents affect the
geometry of the mooring lines comprising the detector.  Meteorological
conditions at the sea surface are critical for deployment and recovery
operations.  The optical background resulting from $^{40}$K decay and
bioluminescence~\cite{article1} will affect trigger rates and track
reconstruction, while bioluminescence bursts can induce dead-time in
the data acquisition.  The parameters which govern the transmission of
Cherenkov light through the sea water to the optical modules are most
important for the design and the performance of the detector; light
absorption largely determines the total number of optical modules required for
efficient detection, while light scattering mainly limits the angular
resolution of the telescope.  Given the objective of operating the
telescope for several years without maintenance, it is mandatory that
the active optical surfaces not be significantly fouled during this
period.
 
This paper reports on a study of optical fouling in the vicinity of 
the ANTARES site ($42^\circ 50'$N $6^\circ 10'$E), which is located 
20~nautical miles (37~km) from Toulon at a depth of 2400~m.  Direct 
measurements have been made of the change in light transmission 
through glass surfaces during immersions of several months.  In order 
to extrapolate to longer periods of time, it is important to 
understand the nature of the fouling, particularly biofouling and 
sedimentation.  Therefore measurements of light transmission were 
complemented by studies of biofouling on glass plates and by a 
detailed study of sedimentation.  Different approaches were used to 
classify and quantify the particle sedimentation; total mass fluxes 
were determined by a time-series collection of samples in a sediment 
trap, particle concentration was measured in water samples taken at 
various depths, and sedimentation rates were calculated from the 
$^{210}$Pb activity in a sea floor core sample.

\section{Fouling and sedimentation measurement 
methods}\label{sec:foulandsed}

Some of the measurements reported here were made using instruments on a 
recoverable mooring line, while others were made on samples obtained 
with the IFREMER deep sea manned submersible Nautile~\cite{nautile}.

\subsection{Mooring line}

The mooring line is anchored to the sea bed by a sinker (dead weight) and held 
vertical by the flotation of buoys.  The line is recovered by the 
remote activation of an acoustic release that disconnects it from the 
sinker.  

Figure~\ref{moor} is a sketch of the mooring line including
information on approximate heights from the sea floor.  The line is
equipped with the following measuring systems (from top to bottom): a
device measuring light transmission between two glass spheres, a
mechanical current-meter\footnote{MC360 from MORS,
www.mors.fr/products/currentmeter/index.html}, a biofilm collection
system, and a sediment trap. These devices are described in the
following sections.

The mooring line was immersed twice at a site ($42^\circ 49'$N 
$6^\circ 10'50''$E) located approximately 1 nautical mile of the 
final site where the ANTARES detector is to be deployed.  The 
first immersion lasted three months in 1997 and the second, eight months 
covering part of 1997 and of 1998.

\subsection{The light transmission measurement system}

The system for measuring light transmission, shown in
figure~\ref{meas}, is housed in two $17''$ pressure resistant glass
spheres similar to those that will house the photomultiplier tubes
(PMT) of the ANTARES detector.  They are mounted on a support frame
with their centres separated by 2.5~m.

The first sphere holds a light source composed of two blue light
emitting diodes\footnote{NSPB500 from NICHIA, www.nichia.co.jp} (LED).
Each LED is monitored by a 5~mm$^2$ PIN photodetector\footnote{OSI5
hybrid PIN photodetector from Centronic, www.centronic.co.uk}
consisting of a photodiode with an operational amplifier.  The two-LED
system is designed to ensure the stability of the light source.  All
of these components are mounted on a holder glued to the inner surface
of the glass sphere.  The distance from the LEDs to the glass is 4~cm;
the size of the illuminated area of the glass is 1~cm in diameter.  A
small mirror on the back of each LED collects the light emitted at
large angles and focuses it on the active area of the monitoring
photodiode.

The second glass sphere contains five photodetectors glued to the 
inner surface of the sphere at various positions.  Each photodetector 
consists of a silicon photodiode with a sensitive area of 1~cm$^2$ and 
an integrated amplifier.  The light flux transmitted to each 
photodiode is measured in order to monitor the effect of fouling on 
the two glass surfaces, in front of the light source and in front of 
the photodiode.

The detector sphere contains the data acquisition board with the 
microprocessor and batteries to power the system.  The microprocessor 
controls the measurement sequence and stores the digitised data (the 
output voltages from the photodiodes).  An acoustic modem allows 
transmission of the data to the sea surface for regular verification 
of the detector status and for intermediate data transfers.

Light transmission measurements were performed twice a day at 0:00 and 
at 12:00 UT. For each photodiode, measurements of the dark current and 
the current produced under illumination with each of the two LEDs, 
were made in a programmed sequence.  Each recorded measurement was the 
average of 10 readings.  The entire sequence lasted 3 minutes.

For the first immersion (deployment January 25, 1997; recovery 
April 21, 1997), the support frame was vertical (as shown in 
figure~\ref{moor}) with the light source on top shining light 
vertically down to the detector sphere. Photodetectors were glued to 
the lower sphere at zenith angles ($\theta$) of $0^\circ$, 20$^\circ$ and 
40$^\circ$. Three photodetectors were placed at 20$^\circ$ on different 
meridians to test for a possible azimuthal ($\phi$) dependence of the fouling. 

For the second immersion (deployment July 12, 1997; recovery March 12, 
1998), the support frame was horizontal and the photodetectors were 
placed at zenith angles ranging from 50$^\circ$ to 90$^\circ$ on a 
single meridian facing the light source, as indicated in 
figure~\ref{meas}.  During this immersion, for an unknown reason, the 
data acquisition stopped on January 31, 1998 and resumed on February 
20, 1998.

\subsection{The biofilm collection system}

The biofilm collection system consists of a 1.2~m long horizontal rod 
supporting 12 cylindrical sample holders.  Six $2.6\times 3.8$~cm$^2$ 
glass plates are mounted on each sample holder as shown in 
figure~\ref{holder}.

During the descent of the mooring line, a cover protects the samples.
A magnesium anode release system triggers the opening of the cover
after a few hours of immersion.  At the end of the exposure, an
externally controlled acoustic release system actuates the closing of
the cover in order to protect the glass plates during the ascent of
the line.  After recovery, all glass slides are immersed in either a
glutaraldehyde or a formaldehyde solution which fixes the biofilm.
They were observed with a scanning electron microscope or stained with
a fluorescent molecule in order to count the total number of bacteria
by epifluorescence microscopy.

\subsection{The sediment trap}\label{sec:trap}

A time-series sediment trap\footnote{Sediment trap PPS5/2 from Technicap,
www.technicap.com} mounted about 100~m above the bottom of the mooring
line collected particles drifting towards the bottom. The trap (2.3~m
high and 1.13~m in diameter) has a 1~m$^2$ collection area and a
baffled conical design (36$^\circ$ opening angle).  The baffle funnels
the sediments into a series of 24 Teflon receiving cups, each having a
volume of 250~ml.  The trap was programmed to collect sediments on a
weekly basis for the first 24 weeks after the immersion (from July 14
to December 29, 1997).

A complete description of the sample processing is given 
in~\cite{Heu}.  Briefly, the receiving cups are filled before 
deployment with a buffered 5\% formaldehyde solution in filtered 
sea water (0.45~$\mu$m) to limit in-situ microbial degradation and to 
reduce contamination by swimmers (organisms entering the trap actively, 
thus introducing an active component in addition to the passive 
settling flux).  After recovery, the cups are stored in the dark at 
2--4$^\circ$C until processed.  The most important step of the 
laboratory processing is the removal of the bulk of swimmers.  Finally 
the remaining samples are de-salted and dried (40$^\circ$C) for 
estimation of mass fluxes and other analyses.

\subsection{Collection of core samples and water samples}

The Nautile submarine, in the period from December 21 to December 24 1998,
collected 6 sediment core samples from the sea floor and 4 water
samples at various depths at the ANTARES site.  To collect each core,
a 10-centimetre diameter 40-centimetre long PVC pipe was thrust
into the sea floor using the manipulator of the Nautile.  The water
samples were collected using plastic Niskin bottles which were
initially held open, then closed using the Nautile manipulator at
different heights from the sea bottom during the ascent of the
submersible.  The cores and the water samples were kept refrigerated
until processed.

\section{Data analysis}

\subsection{Light transmission measurements}

The variation with time of the photodiode current, corrected for 
the dark current contribution, yields the evolution of the light 
transmission.  Since the current depends on the photodiode position 
and sensitivity, the relevant quantity is the variation relative to 
the value measured immediately after immersion.  For each photodiode 
we have checked that these relative variations did not depend (within 
$\pm$~1\% during the first immersion and $\pm$~0.5\% during the second 
immersion) on which of the two LEDs was used.  Therefore the average 
of the measurements obtained for each of the two LEDs was used in the 
analysis.  The resulting relative transmissions as a 
function of time are shown in figures~\ref{fig:foul1}
and~\ref{fig:foul2} for the first and second immersion, respectively.

For both immersions the photodiodes located inside the light source 
sphere gave very stable readings, showing that the light intensity 
from the LEDs was constant to better than $\pm~0.2$\% throughout each 
of the two measurement periods.

Figures~\ref{fig:foul1} and \ref{fig:foul2} show a general trend of 
decreasing fouling with increasing zenith angle on the glass sphere.  
In figure~\ref{fig:foul1}, we observe a very rapid decrease (within a 
few days) in the transmission at the top of the sphere ($\theta = 
0^\circ$), but less change at larger zenith angles.  The transmission 
is seen to recover from time to time in partial correlation with an 
increase in the measured water current velocity (bottom of 
figure~\ref{fig:foul1}).  These observations are consistent with the 
surfaces being fouled by sediments (rather than microbial adhesion and 
growth, see section~\ref{sec:biofouling}) which are more likely to 
stay on horizontal surfaces than on inclined ones, and can be washed 
off by flowing water.  The differences between the three photodiodes 
at zenith angle $20^\circ$ indicate patchiness in the fouling.

Most important for the ANTARES detector, which will operate for several
years without maintenance of the optical modules, is that there is
less fouling at larger zenith angles, as confirmed with the second
immersion, where photodetectors were placed at zenith angles up to
$90^\circ$ (figure~\ref{fig:foul2}). The measured loss in transmission
at the equator of the sphere after 8 months is 2.7\%. The fastest
decrease occurs during the first few days; except for momentary
fluctuations of the order of a percent, the transmission loss 
seems to saturate with time, with a slope of $\sim$0.2\% per month on
average. A linear extrapolation of the $90^\circ$ transmission data
indicates a global loss after 1 year of $\sim$4\%. Since this is the
combination of fouling on two surfaces at $90^\circ$ (the light source
sphere and the detector sphere), the net loss per surface at the
equator after one year is estimated to be half of the total loss, or
$\sim 2$\%.

The ANTARES optical modules, consisting of photomultipliers (PMT)
housed in glass spheres, will be mounted in groups of three, all PMTs
with their axes oriented at $45^\circ$ to the downward vertical
(i.e. at zenith angles of $135^\circ$)~\cite{WEB}. The optically
sensitive region just reaches zenith angles of $90^\circ$
(equatorial).  Therefore we expect the average loss in sensitivity of
optical modules to be small during the several-year operation of
ANTARES. A calibration system is included in the detector design to
monitor the overall light collection efficiency.

\subsection{Sedimentological study}

Results on variations of light transmission through optical surfaces
due to fouling have been complemented by a study of core samples and
measurements of the sedimentation rate.

\subsubsection{Settling particles}

The total mass fluxes have been measured using the sedimentation trap
presented in section~\ref{sec:trap}. They are shown in
figure~\ref{massfl}.  They varied from
19~$\rm mg\cdot m^{-2}\cdot d^{-1}$ (end August) to
352~$\rm mg\cdot m^{-2}\cdot d^{-1}$ (mid October and mid November) with
a clear, apparently seasonal, jump between the two.  In the summer and
early autumn, low amounts of material
($<$100~$\rm mg\cdot m^{-2}\cdot d^{-1}$ typically) were collected.
However, the mass fluxes significantly increased during the second
part of the survey, in the autumn and winter periods (October to
January) as has been previously observed in the Gulf of Lions
(north-western Mediterranean) \cite{Mon1,Mon2}.  This change in the
quantity of collected particles corresponded to a change in the nature
of the material.  The first period was mainly characterised by remains
of biological production (diatoms) whereas the second period was
dominated by lithogenic or detrital material (clays from the
continent).

The low mass fluxes observed during the meteorologically quiet summer
period can be understood from the properties of the site (see map in
figure~\ref{site}):
\begin{itemize}
\item{The location is far from the Rh\^one river system (located 
$\sim$120~km west of Toulon) so the terrestrial influence on the 
ANTARES site should be limited to minor local river systems.} 
\item{The site may be affected on the surface by the 
Liguro-Proven\c{c}al current and by its relatively high biological 
production.  However, the samples were collected deep in the sea 
($>2000$~m), so significant microbial degradation of organic particles 
could occur during the settling time, decreasing the amount of 
particles reaching the sea bottom.}
\end{itemize}

The high mass fluxes measured in some parts of the October to January
period were probably due to the strong meteorological events typical
of this season: flooding rivers and strong winds from the coast.  The
small Mediterranean rivers increase their liquid and solid discharges
considerably during the rainy season.  The location of the site, quite
close to the steep continental slope ($<5$~km from its base), means
that it can be quickly reached by turbid flows originating from the
coast during these episodes.  Moreover the gusty winds can stir up the
sediments on the shelf and contribute to the transfer of particles
towards the abyssal plain.  This explanation is consistent with the
increased amount of clay in the samples collected in this period.  As
shown in figure~\ref{fig:foul2}, however, this period (90 -- 180 days
in figure~\ref{fig:foul2}) was not associated with an increase in the
fouling rate.

\subsubsection{Description of sediment cores}

Four sediment cores have been analysed (the two remaining samples were 
damaged).  All samples gave similar results.  In this paper, we 
discuss the results for a single sample considered as representative 
of the sedimentation patterns occurring at this site on a century time 
scale.

The sediment core shows 4 distinct sections characterised by different 
colours and textures (see figure~\ref{core}).  Starting from the upper 
layer, the first and longest section (0--16 cm) is mainly formed by 
clay sediment and biological material (pteropods), suggesting that particles 
produced in surface sea water strongly contributed to the sedimentation 
processes during the period when this layer was deposited.

The second section (16--21~cm) is dominated by coarse sediments, with a 
smaller fraction of biogenic particles, slowly decreasing from the top 
to the bottom of the section.  The large detrital fraction may be 
interpreted as the consequence of an important resedimentation 
process, such as a turbidity event.  This occurs when sediments 
located at the edge of the continental shelf become unsteady and are 
propelled downslope to the abyssal plain.  From the sedimentation rate 
estimated in the following section, we can infer that the event which 
deposited the second layer occurred over three centuries ago.

The last two sections of the core consist of a thin dark layer on top of a 
substrate formed of grey and brown mud with pteropods. These two 
layers are over three centuries old, so they will not be discussed 
here.

Geochemical analyses were performed to determine water, organic carbon
and carbonate contents.  Total carbon contents were measured by
combustion of dried samples in an analyser\footnote{CS 125 from LECO,
www.leco.com}.  The organic carbon fraction was determined on the
residues remaining after treatment with 99\% pure HCl to remove
inorganic carbon.  All inorganic carbon was assumed to be in the form
of calcium carbonate (CaCO$_3$); the amount was obtained from the
difference between the total carbon content and the organic carbon
content.  Organic carbon and carbonate rates (see figure~\ref{core})
varied within the same ranges as those observed in similar deep sites
in the north-western Mediterranean sea~\cite{Bus}.

\subsubsection{Sedimentation rates from $^{210}$Pb radioactivity}

The sedimentation rate $R$ per unit depth (in $\rm cm\cdot yr^{-1}$) on a
century scale can be determined from the decay profiles of $^{210}$Pb
total activity as a function of depth (or time with $dz=R\,dt$), on 
the basis of the decay rates.

The $^{210}$Pb radionucleide activity of the core was measured by high
resolution alpha spectrometry~\cite{Aba}. Because of a small scale
variability of the water content (and compaction) of the sediment, the
core is cut into 1~cm-thick sections.  All relevant parameters are then 
measured or computed for each of these centimetre-thick layers.  

The $^{210}$Pb radioactivity can be determined as follows:
\begin{equation}
A = A_{ex}(t) + A_s = \frac{F(t)}{r(t)} \exp(-\lambda t) + A_s \;\; ,
\end{equation}
where $A_{ex}$ is the excess activity due to $^{210}$Pb deposited at 
the water-sediment interface, $A_{s}$ is the $^{210}$Pb activity 
supported by the radioactive parents present in the sediment (mostly 
$^{226}$Ra), $\lambda = 0.03114$~yr$^{-1}$ is the $^{210}$Pb decay 
constant, $F$ is the flux of $^{210}$Pb delivered to the sediment (in 
$\rm Bq \cdot cm^{-2}\cdot yr^{-1}$) and $r$ the accumulation rate per 
unit mass of dry sediment (in $\rm g\cdot cm^{-2}\cdot yr^{-1}$).  The 
relation between $r$ and $R$ takes into account the compression of the 
sediment:
\begin{equation}
r = R(z) \times  \rho_{d}\,(1-\Phi(z))\;\; ,
\label{eq:rR} \end{equation}
where $\rho_{d} = 2.55\;\rm g\cdot cm^{-3}$ is the density of dry sediment
and $\Phi$ is the porosity (percentage of the total volume of sediment
that consists of pore spaces) at depth $z$ in the core. Voids in the
wet sediment being filled with water of density $\rho_{\rm water}$,
the porosity is determined from the water mass fraction $f_{\rm
water}=1- m_d/m_w$ with $m_w$ the mass of the wet sediment and $m_d$
the mass of the dry sediment:
\begin{equation}
\Phi = \frac{f_{\rm water}}
       {f_{\rm water}+(1-f_{\rm water})\times\rho_{\rm water}/\rho_d}\;\; .
\end{equation}
Measures of the porosity vary between 0.56 and 0.74 for the various
layers of the core.

The model that best describes the north-western Mediterranean sea
(where the ANTARES site is located) is the Constant Rate of Supply
model where $F(t)$ is assumed to be constant~\cite{CRS}.

The integrated excess $^{210}$Pb activity $I$ below the depth $z$ of 
a given layer is defined as:
\begin{equation}
I = \int_{z}^\infty \rho_{d}\,(1-\Phi(z'))\, A_{ex}(z') \, dz'\;\; .
\end{equation}
The accumulation rate for each layer is then obtained from:
\begin{equation}
r = \frac{\lambda I}{A_{ex}}\;\; ,
\label{eq:r} \end{equation}
and the age of a sediment layer is given by:
\begin{equation}
t = \frac{1}{\lambda}\ln\frac{I_0}{I}\;\; ,
\label{eq:t} \end{equation}
where $I_{0}$ is the integral over the entire sediment core.

Figure~\ref{Pbprof} shows the $^{210}$Pb activity profile, which
decreases rapidly with depth.  A constant contribution $A_{s} =
$~79~$\rm Bq\cdot kg^{-1}$ is estimated from the 7--15~cm section of
the core.  After correcting the $^{210}$Pb activity for this
background, the data within the first 6 cm are used to compute an
average sedimentation rate $\tilde{R}=$~0.052 $\rm cm\cdot yr^{-1}$
(see figure~\ref{fig:sediments}a).

The average accumulation rate $\tilde {r} = 0.040 \rm g \cdot
cm^{-2}\cdot yr^{-1}$ (illustrated in figure~\ref{fig:sediments}b) is
about 8 times higher than the average total mass flux (135~$\rm
mg\cdot m^{-2}\cdot d^{-1}$ or 0.005~$\rm g\cdot cm^{-2}\cdot
yr^{-1}$) determined by the trap.  This discrepancy was also observed
in marine environment studies.  Radakovitch and Heussner suggest three
possible reasons~\cite{Rad99}:
\begin{itemize}
\item{The accumulation rates are overestimated because
of bioturbation. This process decreases the $^{210}$Pb activity at the
water-sediment interface by mixing the surficial sediment layer.}
\item{The sediment trap fluxes were measured on a short period of six
months whereas the average accumulation rate was determined from the
first 6~cm of sediments corresponding to about 150 years. It is
possible that fluxes measured over the 6-month period were lower than
the average sediment accumulation rate over a century.}
\item{Supplementary particles are delivered to the site at
intermediate depths between the trap ($\sim$ 100~m above the sea bed)
and the sediments. The dispersion of suspended particulate matter from
the coastal region to the abyssal plain occurs through nephelometric
structures that may drift along the sea bed without feeding the
sediment trap.}
\end{itemize}

Indeed, the accumulation rate determined above is among the highest 
rates found at similar depths in the North-western Mediterranean sea
\cite{Zuo,Rad}, which is quite surprising considering the lack of
direct particle input from large rivers. This high accumulation rate
may be explained by the short distance from the coast, the steep
continental slope and consequent intensive processes of
re-suspension and transfer of particles towards the abyssal plain. 

An additional core was collected under similar conditions in July 2000 
for analysis of its $^{210}$Pb activity profile.  The excess activity 
is measured in the first 5 centimetres of the core and decreases 
rapidly from 502~$\rm Bq\cdot kg^{-1}$ at surface level to 
1~$\rm Bq\cdot kg^{-1}$ between 4 and 5 centimeters, while the supported 
activity $A_s=71$~$\rm Bq\cdot kg^{-1}$ is measured in the 5--11~cm 
section of the core.  The accumulation rates $r$ measured from the 
various centimetre-thick layers of the core are found to lie in the 
range 0.01--0.02~$\rm g\cdot cm^{-2}\cdot yr^{-1}$ and the sedimentation 
rates $R$ in the range 0.03--0.02~$\rm cm\cdot yr^{-1}$, with an average 
of 0.022~$\rm cm\cdot yr^{-1}$ over the past century.  This value is 
smaller than the one derived from the cores collected in December 1999 
by a factor of 2, suggesting possible local variations of the sediment 
rates, again due to the proximity of the continental slope.  These 
differences remain small however (of the order of a centimetre) on 
the scale of a century.  

\subsection{Water samples}

The suspended particle load in water samples obtained by the
Nautile in December 1998 is shown in figure~\ref{suspload}. 
These data indicate that the suspended load is not constant with 
altitude as would have been expected if the flow of particles was 
purely vertical, indicating some horizontal flux as mentioned earlier.
The interpretation suggested above for the high accumulation rates is
supported by the high suspended particle load (2.5~$\rm mg\cdot l^{-1}$)
measured in water 300~m above the bottom (and to a lesser extent close
to the sea bed), which confirms the presence of nephelometric
structures as vectors of impulsional and seasonal transfer. 

The
reservoir of settling particles, likely originating from the
continental shelf or slope, represents a typical pattern of feeding
deep marine basins.

\subsection{Biofouling}\label{sec:biofouling}

Very little is known about bacterial adhesion on substrates in the
deep sea. Studies at shallower depths have shown that a surface
immersed in an aquatic environment is immediately covered with a
biological slime or biofilm.  The first step, occurring within minutes
of immersion, is the adsorption of organic (carbohydrates, proteins,
humic acids) and inorganic macromolecules already present in the
environment or produced by micro-organisms~\cite{Compere}.  These
adsorbed macromolecules form the primary or conditioning film. This is
an essential step since the resulting modifications of the surface
properties (surface tension, surface free energy, polarity,
wettability) allow subsequent adhesion of micro-organisms such as
bacteria, fungi and algae. The bacterial adhesion itself occurs within
a few hours after immersion. The bacterial attachment to the substrate
is at first reversible, but later becomes irreversible because of the
secretion of extracellular polymers (e.g. acidic exopolysaccharides)
which develop polymeric bridging between the cell and the
substrate. Once the attachment has occurred and if the
physico-chemical conditions at the interface are adequate, bacteria
will grow on the surface as micro-colonies. These colonies and their
extracellular secretions form the biofilm.  The polymers may play an
important role in the loss of light transmissivity of glass spheres.
The bacterial adhesion, the biofilm formation and its growth depend on
different factors such as the environmental physico-chemical
properties (temperature, salinity, dissolved oxygen, organic matter
content, etc.), the substrate nature and micro-roughness, and the
hydrodynamic conditions on the surface.

The density of bacteria after the 3-month or the 8-month exposures at
the ANTARES site, obtained by epifluorescence microscopy, ranges from
10$^4$ to 10$^6$~$\rm bact\cdot cm^{-2}$ (see figure~\ref{bacteries}).
Such low levels are similar to those observed on glass samples exposed
for 1 to 2 weeks in shallow waters at temperatures below 15$^\circ$C
(the temperature measured at the ANTARES site is 13.2$^\circ$C). They
are explained by the low temperature and the poor quality of nutrients
at these depths. Due to the very low bacterial densities, the
systematic error induced by the dispersion from sample to sample masks
any strong dependence with the orientation of the glass plate. There
is only a weak indication that the number of bacteria after the
8-month exposure on glass plates facing upwards is larger than that
observed on glass plates facing sideways or down.

Scanning electron microscopy (SEM) observations of the glass plates confirm
the small total fouling (bacteria and particulates) of the
surfaces. Variations with the plate orientation however is now clearly
visible, as illustrated in the pictures of figure~\ref{fig:SEM}. While
there is almost no deposit on vertical or downward facing plates, some
appear on plates D and more still on plate E. The presence of bacteria
is mostly visible on horizontal plates facing upward (E). The bacteria
observed in the deep sea are smaller than those observed at shallow
depths, and they seem to produce less exopolymeric material.



Light transmission through the bacterial deposits has been measured to
be nearly 100\% for wavelengths ranging from 350 to 850
nm~\cite{BROS}.  The observed attenuation of light transmission should
therefore be attributed to the build-up of loosely adhered particulate
matter.  Indeed, the SEM observations reveal high levels of
particulates, especially on upward facing horizontal glass plates.

The adsorbed material is essentially made of particulates of 
sedimentary origin smaller than 20 $\mu$m.  The washing of the sphere 
surface observed for currents larger than $\sim$10~cm/s points to a 
loose adherence of the particulates on the thin biofilm substrate.  
Most of the particulates were removed from the glass plates during the 
sampling process and transport; on the contrary, biofilm is generally 
very adherent and not easy to remove.
 
\section{Conclusions}

Blue light transmission through glass spheres has been measured over 
several months in the vicinity of the site where the ANTARES neutrino 
telescope will be deployed.  The observed loss of transmissivity 
decreases steadily with increasing zenith angle.  In addition, it 
shows a tendency to saturate with time.  It reached 60\%
on the upper pole of a sphere after a 3-month immersion, but was found
to be only 1.6\% at the equator after 8 months. The loss of light
transmission for a vertical glass surface is estimated to be $\sim 2$\%
after one year.  In order to understand the cause of the 
transmissivity loss, ancillary measurements of sedimentation and 
biofouling were performed.  Despite a fairly large accumulation rate 
at the site, the slow growth of the transparent biofilm substrate 
implies a very loose adhesion of the sediments to the glass surfaces.  
Fouling by deposits of light-absorbing particulates is only 
significant for surfaces facing upwards.
The ANTARES optical modules will be oriented pointing downwards, 
with the minimum zenith angle of the sensitive area of the PMT 
photocathode barely reaching the equator.  Therefore, the loss of 
transmissivity due to the fouling is expected to be small even after several 
years of operation.  

Analysis of the sediment core sample indicates that the most recent 
turbidity event in the site happened more than three centuries ago.
 
\input{ackn}

\begin{figure}[h]
\begin{center}
\caption{The mooring line (figure not to scale), as configured for the
first immersion. The light transmission measuring system was mounted
horizontally for the second immersion. }
\label{moor}
\end{center}
\end{figure}

\begin{figure}[htb]
\begin{center}
\caption{The light transmission measuring system, showing the 
configuration used for the second immersion.}
\label{meas}
\end{center}
\end{figure}

\begin{figure}[htb]
\begin{center}
\caption{Side view of a biofilm sample holder.}
\label{holder}
\end{center}
\end{figure}

\begin{figure}[h]
\begin{center}
\end{center}
\caption {Light transmission as a function of time from the first 
immersion, with the two spheres mounted vertically.  The measurements 
for each of the 5 photodiodes are normalised to unity at immersion, on 
January 25, 1997.  Curves are labeled according to the photodiode 
coordinates on the glass sphere surface (zenith angle $\theta$ and 
azimuthal angle $\phi$).  The current velocity is indicated at 
the bottom of the figure.}
\label{fig:foul1}
\end{figure}

\begin{figure}[h]
\begin{center}
\end{center}
\caption {Light transmission as a function of time from the second 
immersion, with the two spheres mounted horizontally.  The 
measurements for each of the 5 photodiodes are normalised to unity at 
immersion, on July 12, 1997.  Curves are labeled according to the 
photodiode zenith angle $\theta$.}
\label{fig:foul2}
\end{figure}

\begin{figure}
\caption {Total mass fluxes at the ANTARES site over a 6-month
period.}
\label{massfl}
\end{figure}

\begin{figure}[h] \begin{center} 
  \caption{Location
  of the ANTARES site, near the French Mediterranean coast. Contour 
  lines indicate the depth below sea level.}
  \label{site} \end{center}
\end{figure}

\begin{figure}[h] \begin{center} 
  \caption{Description of a 
  core collected at the ANTARES site, with the content profiles of 
  organic carbon and carbonates.  The various shades of grey 
  illustrate the various compositions and textures.}
  \label{core} \end{center}
\end{figure}

\begin{figure}[h]
\caption {$^{210}$Pb raw activity profile.  The shades of grey have 
the same meaning as in figure~\ref{core}.}
\label{Pbprof}
\end{figure}

\begin{figure}[hhh] \centering 
\caption{Sedimentation and accumulation rates over the first 6~cm of
the core (equations \ref{eq:rR} and \ref{eq:r}) as a function of the
epoch at which the sediments deposited (equation \ref{eq:t}).  The
dashed lines indicate the average rates.}\label{fig:sediments}
\end{figure}

\begin{figure}[h]
\caption {Suspended load as a function of altitude from the sea
floor.}
\label{suspload}
\end{figure}

\begin{figure}[h]
\begin{center}
\end{center}
\caption {The bacterial density on the glass plates, at the end of
immersion, as a function of the orientation, for the 2 campaigns.
Points indicate the average values and boxes illustrate the dispersion
from sample to sample.  Orientation labels from A (facing down) to E
(facing up) are defined in Figure~\ref{holder}.}
\label{bacteries}
\end{figure}

\begin{figure}[h]
\begin{center}
\end{center}
\caption {Pictures obtained with the SEM. Left: on a vertical plate,
right: on a horizontal plate. The small arrows in the right-hand
picture show different shapes of free bacteria and bacteria embedded
in exopolymers.}
\label{fig:SEM}
\end{figure}

\end{document}

%% file: SL_fouling.tex

{\bf The ANTARES Collaboration}\\
\vspace{1mm}
\author[lam]{P.~Amram},
\author[dip-genova]{M.~Anghinolfi},
\author[dapnia]{S.~Anvar},
\author[dapnia]{F.E.~Ardellier-Desages},
\author[cppm]{E.~Aslanides},
\author[cppm]{J-J.~Aubert},
\author[dapnia]{R.~Azoulay},
\author[oxford]{D.~Bailey},
\author[cppm]{S.~Basa},
\author[dip-genova]{M.~Battaglieri},
% ...

\vspace{1mm}
{\bf and the CEFREM}\\
\vspace{1mm}
\mbox{\author[cefrem]{J.-C.~Alo\"{\i}si},
\author[cefrem]{Ph.~Kerherv\'e},
\author[cefrem]{A.~Monaco}
}
% ...
\address[cppm]{CPPM -- Centre de Physique des Particules de Marseille, CNRS/IN2P3 Universit\'e de la M\'editerran\'ee Aix-Marseille II, 163 Avenue de Luminy, Case 907, 13288 Marseille Cedex 9, France} 
% ...
\address[dapnia]{DSM/DAPNIA, CEA/Saclay, 91191 Gif Sur Yvette Cedex, France}
\address[dip-bari]{Dipartimento Interateneo di Fisica e Sezione INFN, Via E. Orabona 4, 70126 Bari, Italy} 
\address[dip-bologna]{Dipartimento di Fisica dell'Universit\`a e Sezione INFN, Viale Berti Pichat 6/2, 40127 Bologna, Italy}
\address[dip-catania]{Dipartimento di Fisica ed Astronomia dell'Universit\`a e Sezione INFN, 57 Corso Italia, 95129 Catania, Italy}
\address[dip-genova]{Dipartimento di Fisica dell'Universit\`a e Sezione INFN, Via Dodecaneso 33, 16146 Genova, Italy}
\address[dip-roma]{Dipartimento di Fisica dell'Universit\`a "La Sapienza" e Sezione INFN, P.le Aldo Moro 2, 00185 Roma, Italy}
% ...
\address[lam]{LAM -- Laboratoire d'Astronophysique de Marseille, CNRS/INSU - Universit\'e de Provence Aix-Marseille I, Traverse du Siphon -- Les Trois Lucs, BP 8, 13012 Marseille Cedex 12, France}
% ...
\address[oxford]{University of Oxford, Department of Physics, Nuclear and Astrophysics Laboratory, Keble Road, Oxford OX1 3RH, United Kingdom}
% ...
\address[cefrem]{CEFREM -- Centre de Formation et de Recherche sur l'Environnement Marin, Universit\'e de Perpignan, 66860 Perpignan, France}

\corauth[cor]{Corresponding author: \mbox{Nathalie.Palanque-Delabrouille@cea.fr}}

%% file: ackn.tex

\section*{Acknowledgements}
The authors acknowledge financial support by the funding agencies,
in particular:
Commissariat de l'Energie Atomique, Centre Nationale de la Recherche 
Scientifique, Commission Europ\'eenne (FEDER fund), D\'epartement du Var 
and R\'egion Provence Alps C\^ote d'Azur, City of La Seyne, France;
the Ministerio de Ciencia y Tecnolog\'{\i}a, Spain (FPA2000-1788);
the Instituto Nazionale di Fisica Nucleare, Italy;
the Russian Foundation for Basic Research, grant no. 00-15-96584, Russia;
the foundation for fundamental research on matter FOM and the national
scientific research organization NWO, The Netherlands; 
the Particle Physics and Astronomy Research Council, United Kingdom.
